\documentclass[prl,aps,singlecolumn,superscriptaddress,preprintnumbers,nopacs,floatfix,amsmath,amssymb]{revtex4}

\usepackage{graphicx}
\usepackage{dcolumn}
\usepackage{bm}
\usepackage{amsmath}
\usepackage{graphicx}
\usepackage{amssymb}
\usepackage{mathrsfs}
\usepackage{bbm}
\usepackage{epsfig}

\newcommand{\be}{\begin{eqnarray}}
\newcommand{\ee}{\end{eqnarray}}

\begin{document}

%
%
%
\title{ Solitons and Physics of the Lysogenic to Lytic \\ Transition  in  Enterobacteria Lambda Phage   }

\author{Andrei Krokhotin}
\affiliation{Department of Physics and Astronomy, Uppsala University,
P.O. Box 803, S-75108, Uppsala, Sweden}
\author{Antti J. Niemi}
\affiliation{
Laboratoire de Mathematiques et Physique Theorique
CNRS UMR 6083, F\'ed\'eration Denis Poisson, Universit\'e de Tours,
Parc de Grandmont, F37200, Tours, France}
\affiliation{Department of Physics and Astronomy, Uppsala University,
P.O. Box 803, S-75108, Uppsala, Sweden}

\begin{abstract}
\noindent
The lambda phage is a paradigm temperate bacteriophage. Its lysogenic and lytic life cycles  echo competition between the
DNA binding CI and CRO proteins. Here we address the Physics of this transition in terms 
of  an energy function that portrays the backbone as a multi-soliton configuration. The precision of the individual solitons 
far exceeds the B-factor accuracy of the experimentally determined protein conformations giving
us confidence to conclude  that  three  of the four loops  are each composites of two closely located solitons. 
The only  exception is the repressive DNA binding turn, it is the sole single soliton configuration of the backbone.   
When we compare the solitons with the 
Protein Data Bank we find that the one  preceding the DNA recognition helix is unique to 
the CI protein,   prompting us to conclude that the lysogenic to lytic transition is  due 
to a saddle-node bifurcation involving a  soliton-antisoliton annihilation that removes the first loop.
\end{abstract}


\maketitle

The CI  repressor protein and  its lytic counterpart,  the CRO  protein of the Escherichia coli binding $\lambda$ phage,  
are among the most  extensively studied proteins in molecular biology  \cite{lambda}, \cite{lambda2}. 
They display  a highly intriguing  biological behavior 
by controlling the transitions between the lysogenic  and lytic phases, that has been detailed in numerous
molecular biology textbooks and review articles. 
At a qualitative, mechanistic  level the transition between the lysogenic and lytic 
phases is  quite  well understood \cite{lambda}, \cite{lambda2}. But  at a  quantitative level we 
do not yet understand the physical principle that triggers  the transition. Since lysogeny
is an important example of gene control by repressors, a quantitative Physics based explanation
should have wide  biophysical interest and applicability.

In this Letter we search for a Physics based explanation for the transition between 
lysogenic and lytic  phases  in the $\lambda$ phage. For this we scrutinize the fine structure of the folded
CI repressor protein,
a homo-dimer with 92 residues in each of 
the two monomers \cite{lambda}.  The protein binds  to DNA  with a helix-turn-helix motif  that is  located between the 
residue sites  33-51.  Full crystallographic information  is available in Protein Data Bank (PDB) under 
code 1LMB.

We construct the 1LMB backbone conformation 
explicitly  in terms of  solitons that emerge as classical solutions to the following energy 
function \cite{oma}-\cite{nora},
\begin{equation}
E = - \sum\limits_{i=1}^{N-1}  2\, \kappa_{i+1} \kappa_i  + \sum\limits_{i=1}^N
\left\{ 2 \kappa_i^2 + c\cdot (\kappa_i^2 - m^2)^2\right\}
+ \sum\limits_{i=1}^N \left\{ b \, \kappa_i^2 \tau_i^2 + d \, \tau_i + e \, \tau^2_i 
\right\}
\label{E}
\end{equation}
The summation extends over all residues with $\kappa_i \in [-\pi, \pi] \   {\rm mod}(2\pi)$  the bond 
angle along the lattice that is formed by the  central $C_\alpha$  carbons, 
and  $\tau_i \in [-\pi,\pi] \   {\rm mod}(2\pi)$  the ensuing torsion angle. The parameters  $(c,m,b,d,e)$ are 
all global and specific  to a given motif. Once these angles are known, we can use the discrete Frenet
equation to reconstruct the protein backbone as a piecewise linear polygonal chain.

We emphasize that the energy function (\ref{E}) does not purport to explain the 
details of the atomary level mechanisms that fold the CI protein. Rather, it enables us to examine the 
properties of the  folded CI  in terms of universal physical arguments.

Curiously,  (\ref{E}) has  the functional form of the discretized  
Landau-Ginzburg free energy, that similarly describes the Physics of superconductivity \cite{degennes}:
In a continuum limit the first two terms of (\ref{E}) combine into derivative of curvature that plays the r\^ole of  
Cooper pair density in the Landau-Ginzburg theory.
The third term is the symmetry breaking potential. The fourth term has its origin in spontaneous
symmetry breaking, its presence leads to the notorious Meissner effect in superconductivity \cite{degennes}. 
The fifth term which is absent in the standard Landau-Ginzburg free energy, is the Chern-Simons term that gives 
the protein backbone its chirality. Finally, the last term is like  the Proca mass of a supercurrent. In fact,  in 
(\ref{E})  we have included {\it exactly} all those terms 
that are consistent with general principles of universality and gauge invariance \cite{oma}.

We start by introducing the classical equations of motion for (\ref{E}).  We first eliminate  $\tau_i$ in terms of the bond 
angles

\begin{equation} 
\frac{\partial E}{\partial\tau_i} = 2b\kappa_i^2 \tau_i + 2 e \tau_i + d 
\  = 0 \ \Rightarrow \ \ \tau_i [\kappa_i] = - \frac{1}{2} \cdot \frac{d }{e + b\kappa_i^2} \  \equiv \ - \frac{1}{2} \cdot
\frac{1}{ \frac{e}{d} + \frac{b}{d} \cdot \kappa_i^2 }
\label{Etau}
\end{equation}
Consequently the torsion angles are determined entirely  by the bond angles and the two parameter ratios.
When we substitute (\ref{Etau}) to  the equation  for  $\kappa_i$,  we arrive at 
\begin{equation}
\kappa_{i+1} - 2 \kappa_i + \kappa_{i-1} \ = \ U' [\kappa_i] \kappa_i  \ \equiv\ \frac{dU[\kappa]}{d\kappa_i^2} \ \kappa_i \ \ \ \ (i=1,...,N)
\label{Ekappa}
\end{equation}
(with $\kappa_{0} = \kappa_{N+1} = 0$)
where
\begin{equation}
U[\kappa] = -  \frac{ d}{2} \cdot \tau[\kappa]   - 
2 cm^2 \cdot  \kappa^2 +  c\cdot \kappa^4
\label{U}
\end{equation}
Since the torsion angle depends only on the parameter ratios $\frac{e}{d}$ and $\frac{b}{d}$, if we scale $d$,
$e$ and $b$ equally the profile of $\tau[\kappa]$ remains intact.
In the limit where   $d$ becomes  vanishingly small the first term in (\ref{U}) can then be safely removed,  and
the equation reduces to the ubiquitous spontaneously broken 
discrete nonlinear Schr\"odinger equation \cite{nlse}. Thus, in the $d\to 0$ limit the solution approaches 
the soliton profile of the ensuing continuum equation \cite{maxim},
\begin{equation}
\kappa_i \ = \  (-1)^{r+1}  \frac{ 
m_{r 1}  \cdot e^{ c_{r1} ( i-s_r) } - m_{r2} \cdot e^{ - c_{r2} ( i-s_r)}  }
{e^{ c_{r1} ( i-s_r) } +  e^{ - c_{r2} ( i-s_r)}  }
\label{An1}
\end{equation}
Here $r$ labels the different helix-loop-helix motifs of 1LMB,  and ($c_{r1}, c_{r2}, m_{r1}, m_{r2}, s_r$) are specific to the motif; the
parameter  $s_r$ that is absent in (\ref{E}) specifies the location of the $r$th loop.  The parameters 
($c_{r1}, c_{r2}$) characterize the length of the loop, and ($m_{r1}, m_{r2}$)
together with the ratios ($\frac{e}{d}, \frac{b}{d}$)$_r$ determine the global character of the helices and strands that are adjacent to the
loops. Remarkably this leaves us with  {\it no other} loop specific parameters besides the  $c_r$ that determine the length of the loops, 
and $s_r$  that determine their positions. 

We propose that as such the classical soliton profiles  are duly describing the $C_\alpha$ lattice only in the limit  where thermal
fluctuations vanish. But even near zero temperature the protein remains subject to residual zero-point fluctuations.  It is difficult to estimate
and even harder to accurately calculate the amplitude of these zero-point fluctuations.  As a consequence,  in order to get a realistic order of magnitude
estimate  we have inspected the distribution of the  B-factors that characterize experimental uncertainties, for all  PDB structures where
the crystallographic measurements have been made at temperatures less  than 50K. The result is displayed in Figure 1.
\begin{figure}
  \begin{center}
    \resizebox{12.5cm}{!}{\includegraphics[]{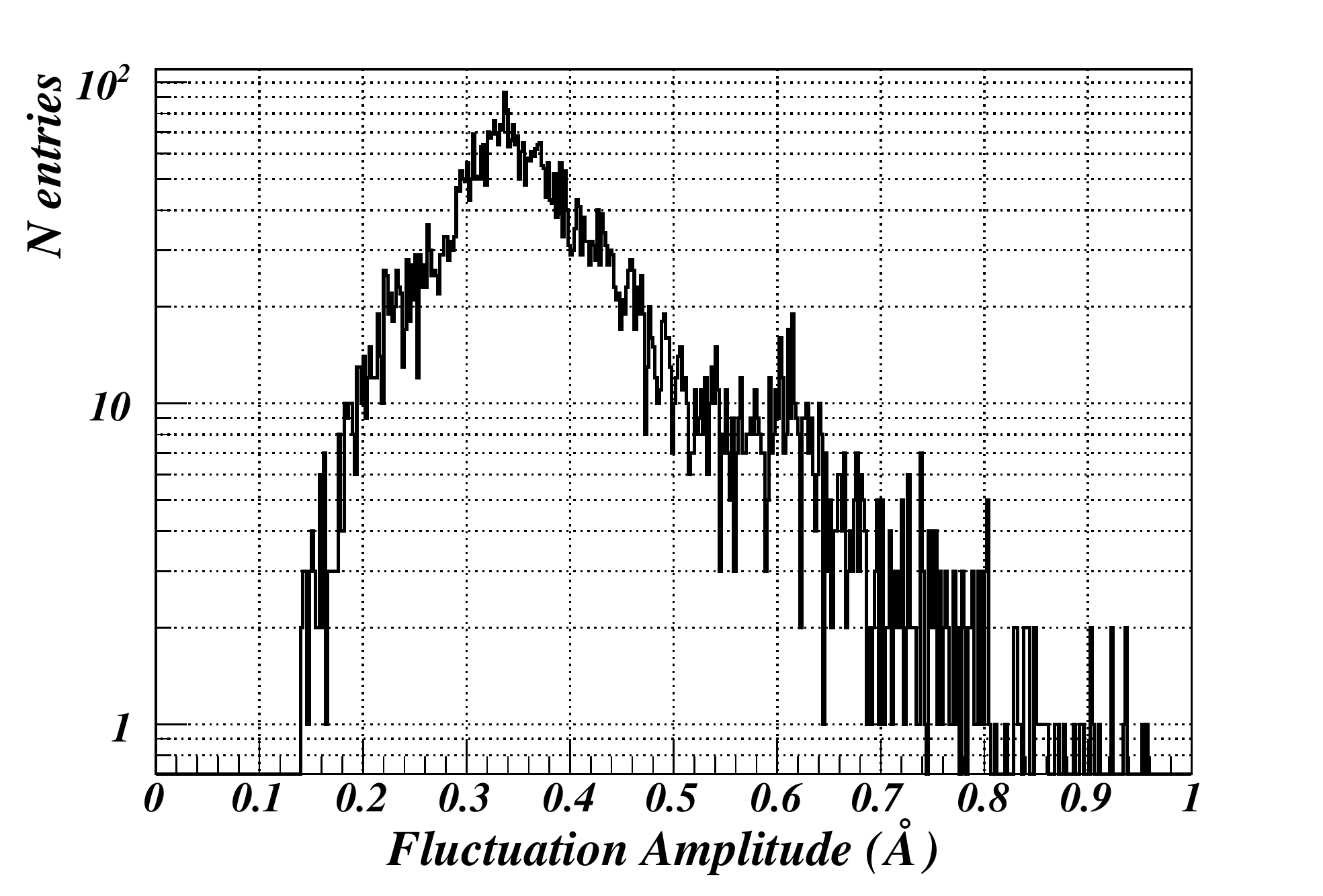}}
    \caption{The number of entries in PDB with temperature below 50K {\it vs.}  Debye-Waller fluctuation distance. }
    \label{fig:simple}
  \end{center}
\end{figure}
From it 
we conclude  that for the $C_\alpha$ carbons the zero point fluctuations have an amplitude somewhere in the vicinity of the lower bound
which is around 0.15 \.A.   Consequently we describe the estimated  range of zero point fluctuations around our classical soliton profiles by dressing  them  
with a tubular dominion that has a radius of 0.15 \.A. 

In Table 1
\begin{table}[tbh]
\begin{center}
\caption{Parameter values for each of the seven solitons in Figure 2. }
\vspace{3mm}
\begin{tabular}{|c|ccccccc|}
\hline
Soliton & $c_1$ & $c_2$ & $m_1$ & $m_2$  & s & e/d & b/d  \\ 
\hline
1& ~ 2.00441 ~ & ~ 1.99595 ~ & ~ 26.65124 ~ & ~ 26.68412 ~   &  ~ 24.50259 & ~ $-9.9921 \cdot 10^{-2}$ ~ & $4.2191 \cdot 10^{-5}$ ~ \\
\hline
2& 2.94889 & 2.95201 & 70.67882 & 70.60369  & \ 30.49642 & $-1.5114 \cdot 10^{-7} $ & $ 1.0662 \cdot 10^{-11} $ \\
\hline
3& 2.89729 & 2.90755 & 39.27387 & 39.22546  &  41.39325 & $-5.3794 \cdot 10^{-7}$ & $7.4566 \cdot 10^{-11}$ \\
\hline
4& 2.97927 & 3.00015 & 1.07948 & 1.52942  & 53.67225 & $  5.1477 \cdot 10^{-7}$ & $ -5.1529 \cdot 10^{-7} $  \\
\hline 
5& 2.96486 & 2.97087 & 26.69087 & 26.25280 &  57.85123 & $-9.62942 \cdot 10^{-8}$ & $1.45097 \cdot 10^{-12}$ \\
\hline
6& 2.94948 & 2.94547 & 20.43071 & 20.38220  &  70.22069 & $-9.27151 \cdot 10^{-7}$ & $3.05202 \cdot 10^{-10}$ \\
\hline
7& 2.89725 & 2.89945 & 89.50870 & 89.55252  &  75.56315 & $-7.13705 \cdot 10^{-7}$ & $1.8457 \cdot 10^{-11}$ \\
\hline 
\end{tabular}
\end{center}
\label{solenoid}
\end{table}
we provide  the parameter values for  (\ref{Etau}) and  (\ref{An1}), computed from the PDB data of 1LMB using a Monte Carlo
fitting algorithm. 
In Figure 2 
\begin{figure}[!hbtp]
  \begin{center}
    \resizebox{12.5cm}{!}{\includegraphics[]{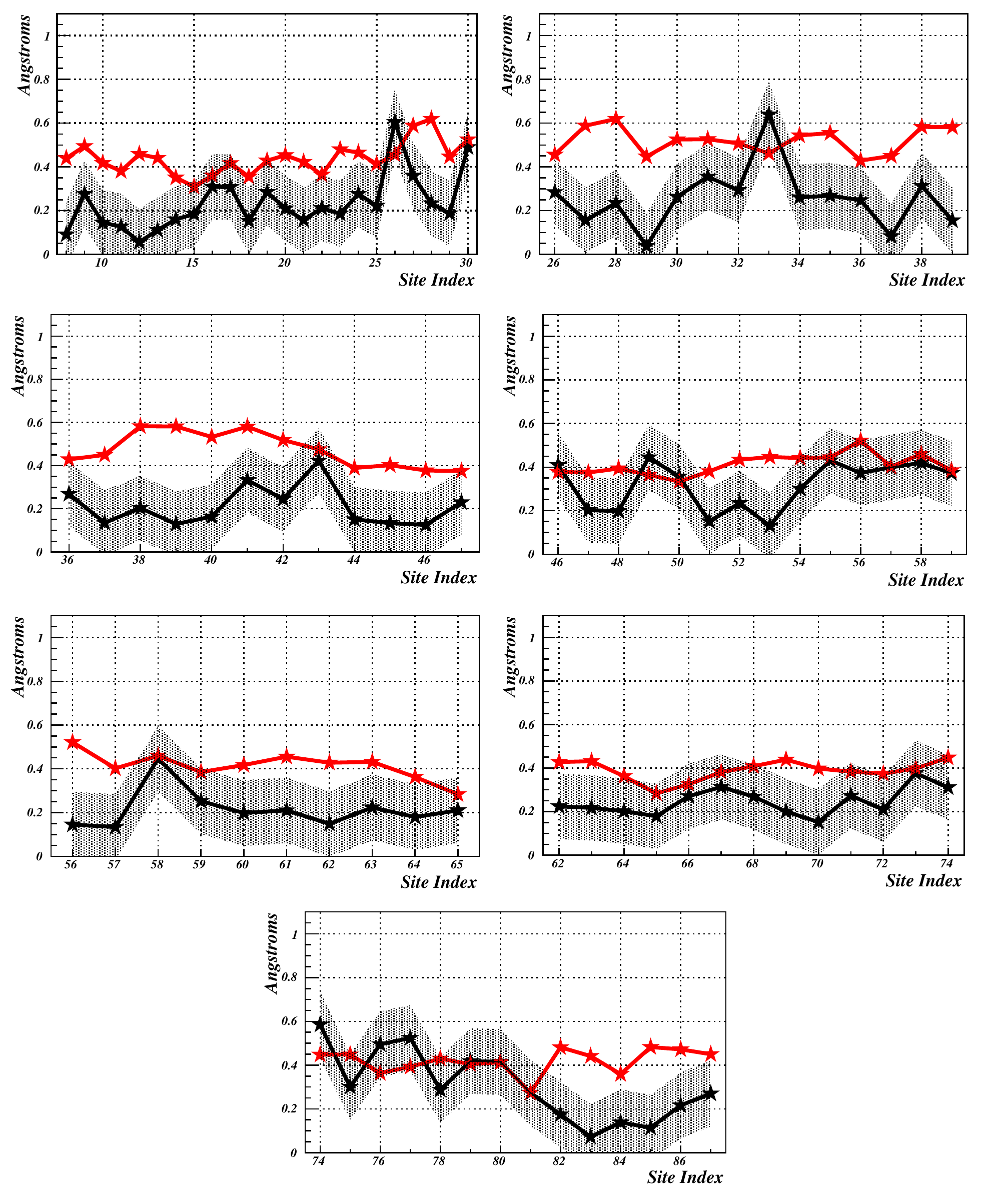}}
    \caption{The seven solitons for the first monomer of 1LMB, with their respective residue numbers. 
    The black line denotes the distance between  soliton 
    and corresponding PDB configuration.  
    The red line denotes the Debye-Waller distance that is computed form the B-factors in PDB.  
    The grey area describes  the estimated 0.15 \.A  zero point fluctuation distance of the soliton.}
    \label{fig:simple}
  \end{center}
\end{figure}
we compare the ensuing solitons with the folded  1LMB: The solitons describe 
the  structural motifs of 1LMB with a precision that  is {\it substantially} better than the experimental 
accuracy determined by B-factors, even when we account for the 0.15 \.A estimate of the solitons zero point fluctuations.

With the aid of our high accuracy solitons we  conclude that in 1LMB there are a total of {\it seven} $\alpha$-helices 
and  {\it one} $\beta$-strand. But  two of the   $\alpha$-helices and the sole $\beta$-strand are 
so short that until now they have been interpreted as parts of loops. They become exposed only by the high accuracy of our construction.
This refinement of the consensus interpretation has important {\it fully testable} repercussions  to  the CI protein that allow 
us to address the Physics of the lysogenic to lytic transition:

The only motif where our soliton picture identifies a loop as a single isolated soliton
is the DNA binding one.  All of the remaining three putative loops consist of a soliton-antisoliton pair, 
with the solitons separated from each other either 
by a {\it very} short $\alpha$-helix in case of the residues  ($23,33$) and ($69,90$), or  by 
a {\it very} short $\beta$-strand in case of  
residues ($51,61$).  This interpretation reveals itself only  when we scrutinize the fine details of the 
($\kappa_i , \tau_i$) spectrum in terms of our solitons.  As an example, in Figure 3
\begin{figure}[!hbtp]
  \begin{center}
    \resizebox{12.cm}{!}{\includegraphics[]{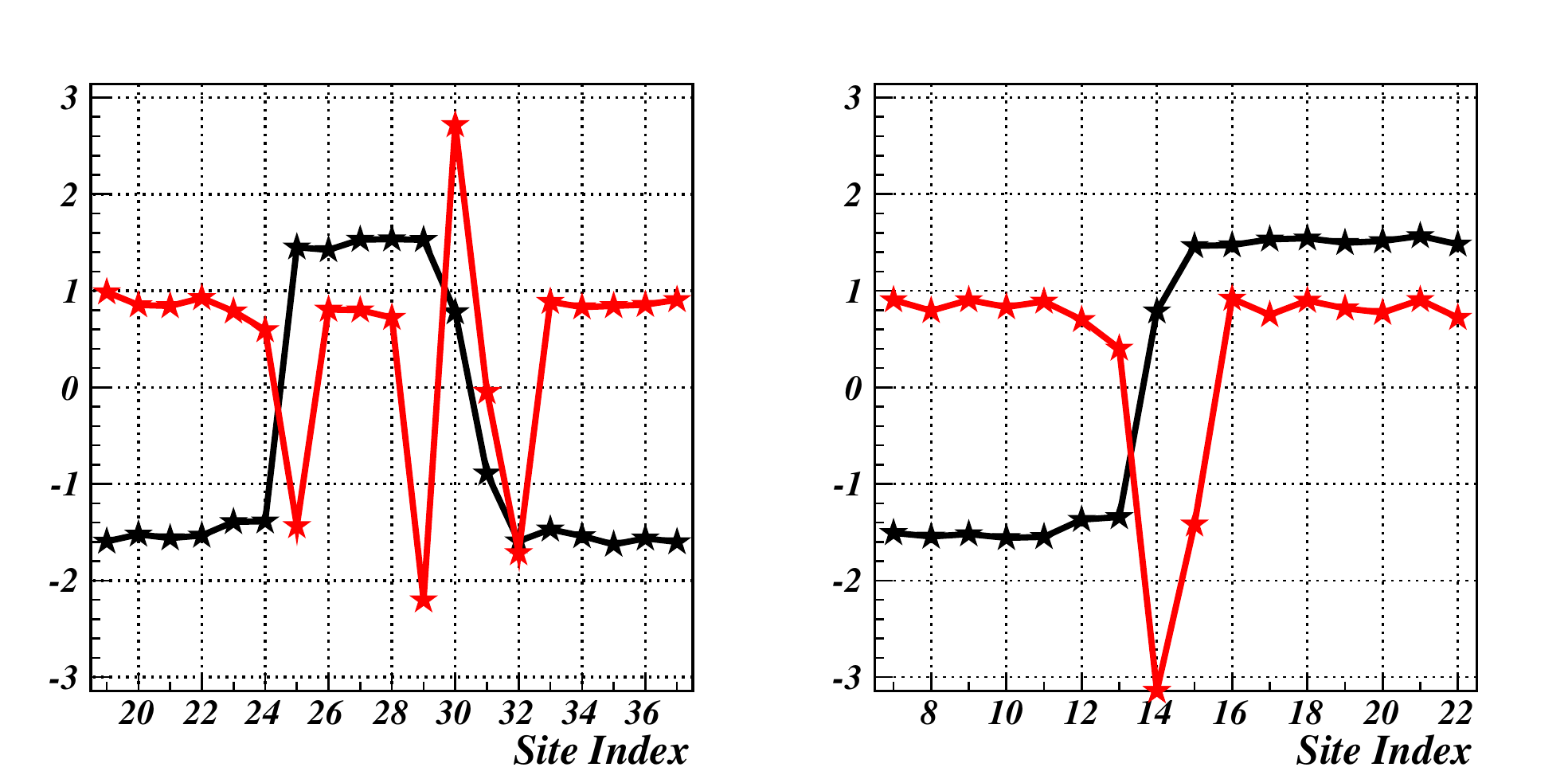}}
    \caption{The resolved ($\kappa_i, \tau_i$) spectrum for the putative first helix-loop-helix of 1LMB (left) and the corresponding structure of in 2OVG (right). 
    The bond angle  $\kappa$ is black, torsion angle  $\tau$ is red. The bond angle
    spectra reveal that in 1LMB the loop is a bound state of two solitons,  while in 2OVG there is only one soliton.}
    \label{fig:simple}
  \end{center}
\end{figure}
we display the putative 
first helix-loop-helix motif and for comparison we display the corresponding structure  in  the CRO 
protein with PDB code 2OVG.  Our  refined interpretation is palpable, in the case of  CI the motif 
is clearly a bound state of two solitons while in the case of  CRO we have a single isolated soliton:

The parameter $s_r$ in (\ref{An1}) determines the center of  soliton {\it i.e.} 
 the position of the  inflection point in the ensuing space curve where  the interpolated 
bond angle in Figure 3 vanishes.
 An isolated  inflection point such as the one in the right hand side of Figure 3 (2OVG)  is topologically stable
in the sense that it can not be created nor removed by any continuous local deformation. For a given finite length curve, 
an individual inflection point {\it i.e.} a soliton can be made or deleted  only by transporting it through one of the end points of the curve.
On the other hand,   a pair of inflection points  {\it i.e.} a soliton-antisoliton pair such as the one in the left hand side of Figure 3 
(1LMB) is not topologically stable but can be created or removed
locally by a saddle-node bifurcation that brings the two  inflection points together. 

A comparison between the CI and CRO soliton
profiles  in Figure 3 then proposes the following {\it experimentally  fully testable} mechanism for the lysogenic-lytic transition: 
Under lysogenic conditions where the CI protein prevails, the soliton-antisoliton pairs of the CI protein that are located 
immediately prior and after the DNA binding domain are relatively stable. But when there is a change in the 
environmental conditions that excites phonon fluctuations along the protein chain, such as raise in temperature or
UV radiation, either of these  soliton-antisoliton pairs can discharge by a saddle-node bifurcation.
This  bifurcation disturbs the structure of the immediately adjacent DNA binding motif  to the extent that 
the protein looses its capability to maintain the lysogenic phase. 
Since each of the corresponding motifs in the CRO protein are  topologically stable single solitons they are insensitive
to local phonon excitations,  and the lytic phase takes over. 

We note that  the shoulder of the short $\alpha$-helix  in the last loop is anchored by the presence of a proline at site 78.
Consequently in the first approximation we can safely exclude a  bifurcation instability from occurring in
the putative helix-loop-helix motif between the residues ($69,90$).

We still need to conclude which of the two motifs of CI that are adjacent to the DNA binding domain  
is the one that looses its  stability in the proposed bifurcation transition. Unfortunately, it appears that a full answer must wait until 
computational methods have reached sufficient maturity \cite{nat}. However, to provide the probable answer
we have performed a statistical analysis on the occurrence of our seven solitons in all PDB proteins. 
In Table 2 
\begin{table}[!htb]
\begin{center}
\caption{ The soliton sites used in searching for matching structures in PDB, together with the number of matches. 
The search is limited to those x-ray structures that have a resolution better than 2.0 \.A and a match is a configuration that 
deviate less than 0.5 \.A in total RMSD distance from the soliton.} 
\vspace{3mm}
\begin{tabular}{|c|ccccccc|}
\hline
Soliton & 1 & 2 & 3 & 4 & 5 & 6 & 7 \\
\hline
Sites   & ~(20,28)  ~& ~(27,36)  ~& ~(36,46)  ~& ~(50,58)  ~  & ~ (55,63)  ~&~ (66,75) ~ & ~(74,82)~  \\
\hline 
Matches  & 9601  & 4  & 810  & 159    & 1552  & 1342 & 406  \\ 
\hline 
\end{tabular}
\end{center}
\label{table2}
\end{table}
we list the number of matches that each of these soliton has when we search PDB for 
configurations that deviate from the given soliton by an overall  RMSD distance  less than 0.5 \.A.  
We have chosen this  cut-off value since it is representative of  the Debye-Waller fluctuation distance 
in the experimental 1LMB data. The {\it remarkable} observation  is that for the second 
soliton in the loop preceding the DNA recognition helix,  the {\it only} matching structures are located
in the different PDB entries 
of the $\lambda$ phage CI protein itself. This absence of the second soliton in PDB {\it strongly} proposes 
that the ensuing loop must be unstable when the protein is in any other {\it in vivo} environment. Thus 
the most probable source of the lysogenic to lytic transition is  the saddle-node  bifurcation that takes place in the
first loop and makes its soliton-antisoliton pair to annihilate each other.  The bifurcation causes the ensuing structure to 
act like a crowbar that lifts the recognition helix from its place in the DNA groove.
We note that {\it as such}, it is obvious from (\ref{E}) that  in isolation the first soliton-antisoliton pair has an 
energy which is higher than that of the ground state {\it i.e.} an $\alpha$-helix. But a more detailed molecular dynamics 
simulation needs to be performed to confirm our proposal.

All of the other solitons are ubiquitous in PDB  and  except for the turn that participates directly 
to the regulatory process  their biophysical r\^ole  remains to be clarified.

Finally, our soliton interpretation reveals the  following {\it fully testable} pattern for the  folding pathways of 1LMB:
The functionally pertinent DNA binding loop  is an isolated soliton, while all the remaining three loop structures 
consist of a soliton-antisoliton pair. Since an isolated soliton is topologically stable while soliton-antisoliton pairs are
not, the DNA binding loop must be created very early, presumably during translation. The initial configuration for the 
folding process is then a single soliton state.  During the folding process the remaining three motifs are created by
local phononic fluctuations, as soliton-antisoliton pairs. Due to the presence of the proline at site 78, the ensuing motif probably 
emerges very early.

\vskip 0.5cm
A.N.  thanks H. Frauenfelder and  G. Petsko  for communications and J. \.Aqvist for discussions. 

\vskip 2.6cm

\end{document}